%% file: main.tex
\documentclass[conference]{IEEEtran}
\IEEEoverridecommandlockouts
\usepackage{cite}
\usepackage{amsmath,amssymb,amsfonts}
\usepackage{algorithmic}
\usepackage{graphicx}
\usepackage{textcomp}
\usepackage{xcolor}
\usepackage{booktabs}
\usepackage[acronym]{glossaries}
\usepackage{todonotes}
\usepackage{url}
\usepackage{siunitx}
\usepackage{enumitem}
\def\BibTeX{{\rm B\kern-.05em{\sc i\kern-.025em b}\kern-.08em
    T\kern-.1667em\lower.7ex\hbox{E}\kern-.125emX}}

\usepackage{xurl}

\usepackage[breaklinks=true]{hyperref}
\hypersetup{pdfstartview=FitH,
  pdfpagelayout=SinglePage,
  bookmarksnumbered=true,
  bookmarksopen=true,
  colorlinks=true,
  citecolor=blue,
  linkcolor=blue
}

\begin{document}

\newacronym{RSRP}{RSRP}{Reference Signal Received Power}
\newacronym{UE}{UE}{User Equipment}
\newacronym{SNR}{SNR}{Signal-to-noise ratio}
\newacronym{PRI}{PRI}{Parameter-Refresh Interval}
\newacronym{SDT}{SDT}{UE State Dwell Time}
\newacronym{UDR}{UDR}{Update-to-Dwell Ratio}
\newacronym{MLP}{MLP}{Multi-Layer Perceptron}
\newacronym{POMDP}{POMDP}{Partially Observable Markov Decision Process}
\newacronym{SINR}{SINR}{Signal-to-Interference-plus-Noise Ratio}
\newacronym{RSSI}{RSSI}{Received Signal Strength Indicator}
\newacronym{SIB}{SIB}{System Information Block}
\newacronym{SE}{SE}{Spectral Efficiency}

\newcommand{\namex}{CellPilot}

\newcommand{\mi}[1] {\todo[color=blue!20,inline]{Marvin: #1}}
\newcommand{\lin}[1] {\todo[color=green!20,inline]{Lin: #1}}
\newcommand{\ramin}[1] {\todo[color=red!20,inline]{Ramin: #1}}
\newcommand{\guto}[1] {\todo[color=yellow!20,inline]{Guto: #1}}
\newcommand{\insight}[1]{\hspace{.25em}$\blacktriangleright$\hspace{5pt}#1}
\newcommand{\update}[1]{#1\textcolor{red}{$\blacktriangleright$update value}}

\title{Cells on Autopilot: Adaptive Cell (Re)Selection via Reinforcement Learning}

\author{
\IEEEauthorblockN{Marvin Illian\IEEEauthorrefmark{1}, Ramin Khalili\IEEEauthorrefmark{2}, Antonio A. de A. Rocha\IEEEauthorrefmark{3}, Lin Wang\IEEEauthorrefmark{1}}\\
\IEEEauthorblockA{\IEEEauthorrefmark{1}Paderborn University, Germany \quad
\IEEEauthorrefmark{2}Huawei Technologies, Germany \quad
\IEEEauthorrefmark{3}Fluminense Federal University, Brazil}
}

\maketitle

\input{text/abstract}

\input{text/introduction}
\input{text/model}
\input{text/algorithm}

\input{text/generalization}
\input{text/evaluation}

\input{text/discussion}

\input{text/relatedwork}

\input{text/conclusion}

\bibliographystyle{IEEEtran}
\bibliography{refs}

\end{document}

%% file: text/abstract.tex
\begin{abstract}
The widespread deployment of 5G networks, together with the coexistence of 4G/LTE networks, provides mobile devices a diverse set of candidate cells to connect to. However, associating mobile devices to cells to maximize overall network performance, a.k.a. cell (re)selection, remains a key challenge for mobile operators. Today, cell (re)selection parameters are typically configured manually based on operator experience and rarely adapted to dynamic network conditions. In this work, we ask: Can an agent automatically learn and adapt cell (re)selection parameters to consistently improve network performance? We present a reinforcement learning (RL)-based framework called \namex{} that adaptively tunes cell (re)selection parameters by learning spatiotemporal patterns of mobile network dynamics. Our study with real-world data demonstrates that even a lightweight RL agent can outperform conventional heuristic reconfigurations by up to 167\%, while generalizing effectively across different network scenarios. These results indicate that data-driven approaches can significantly improve cell (re)selection configurations and enhance mobile network performance. 
\end{abstract}

%% file: text/introduction.tex
\section{Introduction}
The widespread adoption of 5G networks, coupled with the continued operation of 4G/LTE infrastructure, has significantly expanded the range of connectivity options for mobile devices~\cite{bna5GSA,fccSpectrum2022}. This coexistence creates a heterogeneous environment where multiple cells\footnote{We define a cell as a sector of a base station that operates on a specific frequency band, covering a specific area of the network.}---potentially operating on different frequency bands---are simultaneously accessible to \glspl{UE}~\cite{2024-hotmobile-bandswitching}. Such an environment provides unprecedented flexibility in cell (re)selection in idle mode\footnote{
Idle-mode cell selection plays a critical role in multi-carrier network deployments, enabling efficient load balancing among users without complex signaling. It  reduces overhead and errors associated with connected-mode load balancing, ensuring smoother user transitions to active mode~\cite{2016-wcnc-layer}.
}, enabling mobile devices to dynamically choose between cells based on signal strength and other mobility-related factors~\cite{2021-icnp-cellselection}. While mobile operators gain the opportunity to optimize connectivity decisions to improve network efficiency and user experience, this also introduces significantly higher complexity in designing effective cell (re)selection strategies. 

\begin{figure}[!t]
    \centering
    \includegraphics[width=0.9\linewidth]{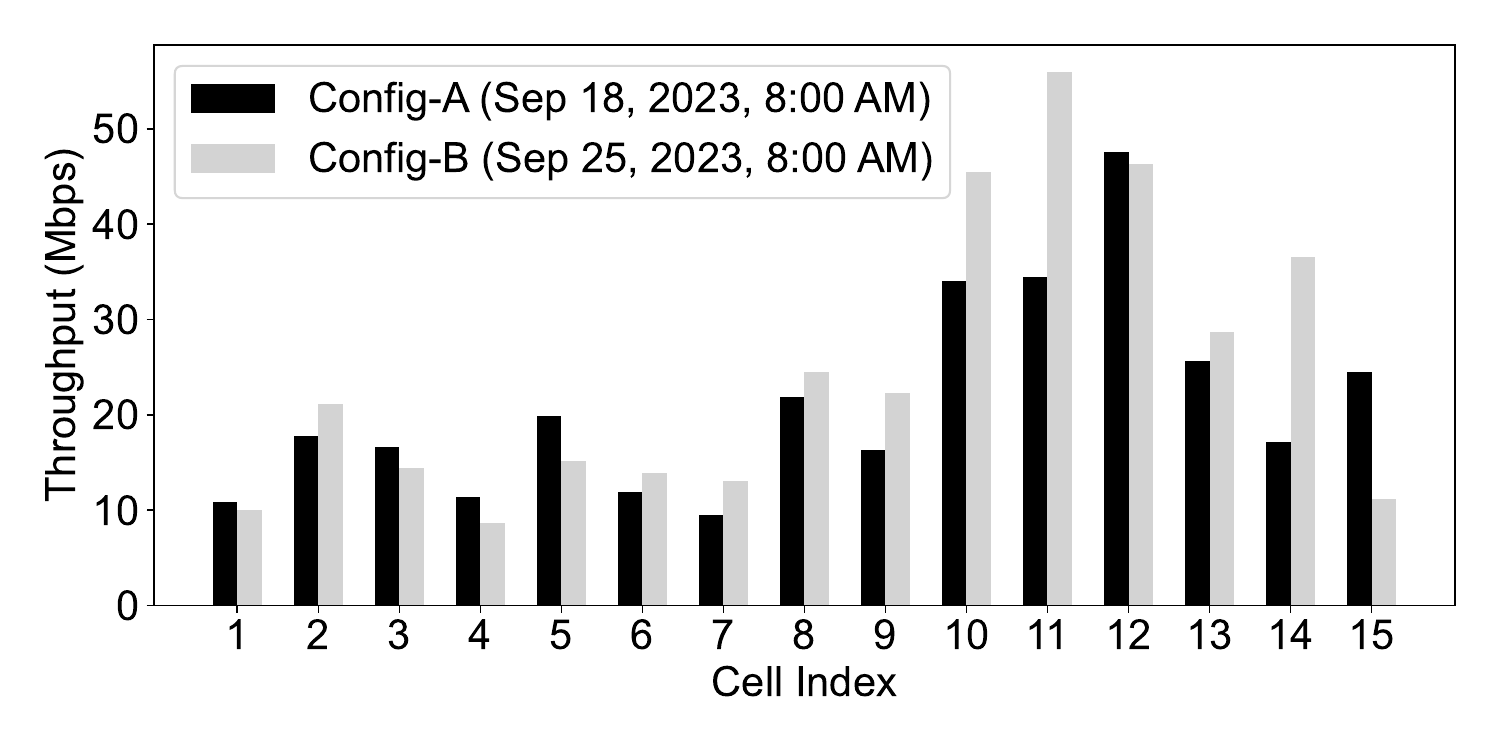}
    \caption{Per-\gls{UE} downlink throughput comparison across 15 cells in a city in Brazil, under two different parameter configurations for cell (re)selection.}
    \label{fig:configs}
\end{figure}

Despite the offered flexibility, most mobile operators still rely on fixed parameter configurations for cell (re)selection.
In practice, operators lack a standardized methodology for these configurations which are typically dervived experience-based by network engineers and refined iteratively through trail and error.
These parameters, including various thresholds and priorities, are typically determined through offline planning and are seldom adapted in real time to reflect changing network conditions~\cite{2020-hotmobile-cellselection,2021-icnp-cellselection}. While such static configurations simplify network management, they often fail to exploit the full potential of heterogeneous mobile network deployments, leading to suboptimal network utilization and uneven load distribution. Through a case study conducted with one of the three largest mobile operators in Brazil, we show in Figure~\ref{fig:configs} that under roughly the same network loads (Config-B has 9\% less \glspl{UE}) but different parameter configurations, the per-\gls{UE} downlink throughput in a cell can vary up to two times. This highlights the substantial impact of cell (re)selection parameter reconfiguration on network performance and hence underlines the need for more adaptive approaches capable of dynamically adjusting cell (re)selection strategies to optimize network performance.

Adaptive parameter configuration for cell (re)selection poses significant challenges due to multiple reasons. First, to be compatible with existing network infrastructure and operations with minimal changes, cell (re)selection decisions cannot be made explicitly and should be the outcome of parameter configurations implicitly. The parameters form a huge and high-dimensional space, integrating several thresholds and priorities that exhibit non-linear interactions. Moreover, the parameters concentrate mostly on signal strength and are not directly performance-oriented. Hence, it is difficult to associate a particular parameter configuration to a performance level, making parameter reconfiguration intrinsically a hard problem. 

Second, mobile networks are highly dynamic in nature, both spatially and temporally. Variations in user mobility patterns, traffic demand fluctuations, and interference levels make it difficult to determine optimal parameters in real time. In a highly heterogeneous network, the impact of changing one parameter can cascade through the network, affecting network utilization and load distribution. Traditional profiling- and experience-based optimizations struggle to cope with this scale and complexity and typically require extensive manual efforts~\cite{2021-icnp-cellselection}. We conjecture that a learning-based approach is more suitable. A learned agent can autonomously explore the high-dimensional parameter space, learning policies that dynamically adapt to changing network conditions and improving network performance and efficiency. 

In light of the above challenge, we propose \namex{}, a neural adaptive cell (re)selection framework for mobile networks. \namex{} adopts reinforcement learning (RL) to intelligently optimize the parameters for cell (re)selection under dynamic user connectivity and mobility. The key innovation of \namex{} lies in its state and reward design, where \gls{UE} status and network load conditions, both historical and current ones, are encoded to capture the dynamic network conditions. The reward function is carefully crafted to incorporate multiple objectives. Furthermore, \namex{} introduces a series of optimization techniques to enhance RL agent generalizability, ensuring robust performance under diverse traffic patterns and across network scenarios without retraining. The combination of principled RL modeling and taillored optimizations enables \namex{} to outperform the conventional heuristic-based approach and adapt seamlessly to evolving network environments. 

Overall, we make the following contributions: After introducing the cell (re)selection problem (\S~\ref{sec:problem-modeling}), we

\begin{itemize}
    \item propose \namex{}, a learning-based framework for adaptive tuning for cell (re)selection, leveraging reinforcement learning to capture the dynamic network environment spatiotemporally (\S~\ref{sec:algo}). 
    \item introduce, adapt, and integrate several optimization techniques to improve the performance and generalizability of \namex{} (\S~\ref{sec:generalization}).
    \item build a simulator with real-world network data and comprehensively evaluate \namex{} under varying conditions (\S~\ref{sec:eval}). Our results show that \namex{} achieves up to 167\% performance improvements. \namex{} also shows impressive generalizability, despite being lightweight. 
\end{itemize}
We discuss the limitations of \namex{} in \S~\ref{sec:discussion}, contrast \namex{} with related works in \S~\ref{sec:relatedwork} and conclude in \S~\ref{sec:conclusion}.

%% file: text/model.tex
\section{The Cell (Re)Selection Problem}
\label{sec:problem-modeling}

In this section, we provide the background information and formally describe the cell (re)selection problem.

\subsection{Cell (Re)Selection Process}

The cell selection process in a mobile network~\cite{2016-wcnc-layer, 3GPP2018-br, 3GPP2019-dn} is detailed as follows.
The process is identical for 4G/LTE and 5G, with only minor parameter differences.
Each cell is powered by a base station operating on a specific carrier frequency and is identified by a unique cell ID~\cite{3GPP2017-rl}. When a \gls{UE} powers up and enters a network, it performs cell selection and the selection criteria are based on \gls{RSRP} and broadcast parameters. 

Before a \gls{UE} considers any cell for camping, it verifies if the signal fulfills the suitability condition.
This requirement ensures that the UE can maintain adequate communication with the selected cell and is expressed as
\begin{equation}
    S^{C_i}_{\mathrm{rxlev},n} = \gamma^{C_i}_{n} - Q^{C_i}_{\mathrm{rxlevmin}} > 0
    \label{eq:suitability-condition}
\end{equation}
where $\gamma^{C_i}_{n}$ represents \gls{RSSI} for \gls{UE} $n$ on cell $C_i$, $Q^{C_i}_{\mathrm{rxlevmin}}$ denotes the minimum required \gls{RSSI} for cell $C_i$ as broadcast in the \gls{SIB}, and $S^{C_i}_{\mathrm{rxlev},n}$ defines the cell (re)selection received level for \gls{UE} $n$ on cell $C_i$.

Once the suitability condition is satisfied, the \gls{UE} scans for all available cells, selects the most appropriate one based on network-defined criteria and camps on it.
The \gls{UE} then monitors the \glspl{SIB} to stay informed about network configuration changes and neighboring cells.
The network then guides idle-mode (re)selection through two complementary strategies: hierarchical cell (re)selection and equal priority cell (re)selection, as detailed in the following.

\subsection{Hierarchical Cell (Re)Selection Strategy}

In a hierarchical cell (re)selection strategy, the network assigns different priorities to cells.
This allows the network to steer \glspl{UE} towards a cell that provides best possible service by avoiding highly-congested cells even if they have a stronger signal.
When the serving cell $C_i$ has a lower (re)selections priority than a neighboring cell $C_j$, the reselection criteria focus on the \gls{RSRP} of the neighboring cell.
The reselection process is initiated when the \gls{RSRP} of $C_j$ exceeds a threshold $T^{C_i \rightarrow C_j}_{\mathrm{XHigh}}$ for the duration of the reselection timer $T_{\mathrm{reselection}}$ to avoid frequent reselections if brief fluctuations occur:
\begin{equation}
\gamma^{C_j}_{n} > T_{\mathrm{XHigh}}^{C_i \rightarrow C_j} \; \text{for duration} \; \; T_{\mathrm{resel}}.
\end{equation}
If multiple cells meet the criterion, the \gls{UE} will reselect the highest-ranked cell among the highest priority frequencies.

Reselection to a neighboring cell with a lower priority involves multiple criteria, and works against the network's preferred hierarchy.
For inter-frequency reselection, the process is initiated when the serving cell's \gls{RSRP} $S_{\mathrm{rxlev}}^{C_i}$ falls below a threshold $S^{\mathrm{Inter}}$:
\begin{equation}
    S^{C_i}_{\mathrm{rxlev}} < S^{\mathrm{Inter}}.
\end{equation}
This process is performed analogously for intra-frequency reselection using $S^{\mathrm{Intra}}$.
Subsequently, reselection to a lower priority cell is triggered if both the serving cell's \gls{RSRP} $\gamma^{C_i}_{n}$ and the neighboring cell's \gls{RSRP} $\gamma^{C_j}_{n}$ meet the following conditions for the duration of the reselection timer $T_{\mathrm{resel}}$:
\begin{equation}
\begin{aligned}
    \gamma^{C_i}_{n} < T^{C_i}_{\mathrm{SLow}} \; \text{for duration} \; \; T_{\mathrm{resel}}, \\
    \gamma^{C_j}_{n} > T^{C_i \rightarrow C_j}_{\mathrm{XLow}} \; \text{for duration} \; \; T_{\mathrm{resel}}.
\end{aligned}
\end{equation}

\subsection{Equal Priority Cell (Re)Selection Strategy}

Equal priority cell (re)selection is employed when the \gls{RSRP} difference between cells is typically less than \SI{3}{\deci\bel}.
In this case, the absolute \gls{RSRP} levels become not as critical and the focus shifts to relative differences.
This strategy directs users to cells with stronger \gls{RSRP} while allowing network operators to fine-tune the traffic distribution.
The offset parameter $Q^{C_i \rightarrow C_j}_{\mathrm{Offset_{freq}}}$ can be dynamically adjusted by the network to influence the \gls{UE} reselection process.
By increasing this offset, the network can discourage \glspl{UE} from selection to certain neighboring cells, effectively balancing traffic according to prevailing network conditions and meeting congestion or quality of service targets.

The reselection in equal priority applies ranks to the serving and neighboring cells based on their \gls{RSRP} and the offset parameter.
The rank of the neighboring cell is defined as
\begin{equation}
    R_{\mathrm{Nbr}} = \gamma^{C_j}_{n} - Q^{C_i \rightarrow C_j}_{\mathrm{Offset_{freq}}}
\end{equation}
To avoid frequent reselections, the rank of the serving cell is adjusted with a hysteresis value $Q^{C_i}_{\mathrm{Hyst}}$ and is defined as
\begin{equation}
    R_{\mathrm{Ser}} = \gamma^{C_i}_{n} + Q^{C_i}_{\mathrm{Hyst}}
\end{equation}
Again, a timer $T_{\mathrm{reselection}}$ ensures that the reselection decision is robust to brief fluctuations in \gls{RSRP}.
\begin{equation}
    R_{\mathrm{Nbr}} > R_{\mathrm{Ser}} \;\;\text{for duration} \;\; T_{\mathrm{resel}}
\end{equation}

\subsection{Conventional Heuristic Baseline}

\begin{table}[!t]
\caption{Conventional Heuristic Reference Values}
\label{tab:heuristic-reference}
\centering
\begin{tabular}{@{}p{0.055\linewidth}
                >{\centering\arraybackslash}p{0.065\linewidth}
                >{\centering\arraybackslash}p{0.11\linewidth}
                >{\centering\arraybackslash}p{0.12\linewidth}
                >{\centering\arraybackslash}p{0.12\linewidth}
                >{\centering\arraybackslash}p{0.12\linewidth}
                >{\centering\arraybackslash}p{0.12\linewidth}@{}}
\toprule
\textbf{Symb.} &
$Q^{C_i}_{\mathrm{Hyst}}$ &
$Q^{C_i \rightarrow C_j}_{\mathrm{Offset_{freq}}}$ &
$Q^{C_i}_{\mathrm{rxlevmin}}$ &
$T^{C_i \rightarrow C_j}_{\mathrm{XHigh}}$ &
$T^{C_i \rightarrow C_j}_{\mathrm{XLow}}$ &
$T^{C_i}_{\mathrm{SLow}}$ \\
\midrule
\textbf{Val.} &
3 dB & 14 dB & -60 dBm & -56 dBm & -58 dBm & -54 dBm\\
\bottomrule
\end{tabular}
\end{table}

As a reference point for performance comparison, we use the network operator's conventional heuristic parameter set deployed in Petr\'{o}polis, which is also the evaluation region in our study.
This configuration is specifically tuned for the Petr\'{o}polis area and corresponds to the better-performing Config-B in Figure~\ref{fig:configs}, ensuring that our RL approach is not compared against a weak or untuned baseline.
The exact parameter values are listed in Table~\ref{tab:heuristic-reference}.

\subsection{Challenges in Manual Parameter Tuning}

\begin{table}[!t]
\caption{Alternative Heuristic Configuration (Config-A)}
\label{tab:config-a}
\centering
\begin{tabular}{@{}p{0.055\linewidth}
                >{\centering\arraybackslash}p{0.065\linewidth}
                >{\centering\arraybackslash}p{0.11\linewidth}
                >{\centering\arraybackslash}p{0.12\linewidth}
                >{\centering\arraybackslash}p{0.12\linewidth}
                >{\centering\arraybackslash}p{0.12\linewidth}
                >{\centering\arraybackslash}p{0.12\linewidth}@{}}
\toprule
\textbf{Symb.} &
$Q^{C_i}_{\mathrm{Hyst}}$ &
$Q^{C_i \rightarrow C_j}_{\mathrm{Offset_{freq}}}$ &
$Q^{C_i}_{\mathrm{rxlevmin}}$ &
$T^{C_i \rightarrow C_j}_{\mathrm{XHigh}}$ &
$T^{C_i \rightarrow C_j}_{\mathrm{XLow}}$ &
$T^{C_i}_{\mathrm{SLow}}$ \\
\midrule
\textbf{Val.} &
3 dB & 20 dB & -60 dBm & -58 dBm & -60 dBm & -58 dBm\\
\bottomrule
\end{tabular}
\end{table}

\begin{figure}[!t]
    \centering
    \includegraphics[width=0.9\linewidth]{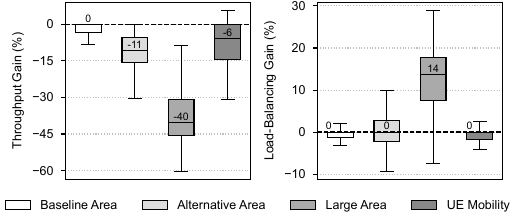}
    \caption{Throughput and load-balancing gains achieved by Config-A across different network scenarios, normalized to Config-B (zero line).}
    \label{fig:config-a-gain}
\end{figure}

To further illustrate the challenges of manual parameter tuning, Figure~\ref{fig:config-a-gain} compares the performance of Config-A against the better-performing Config-B across different simulated network scenarios.
The simulation results show the same qualitative trend as the operator KPIs in Figure~\ref{fig:configs} suggests: Config-A consistently underperforms Config-B in terms of throughput.
Moreover, Figure~\ref{fig:config-a-gain} reveals conflicting optimization objectives.
In the large-area scenario, Config-A slightly improves load balancing but underperforms in throughput.
Trade-offs like these illustrate the difficulty of manually deriving parameter configurations that generalize well, underscoring the case for automated, learning-based parameter optimization.

%% file: text/algorithm.tex
\section{\namex{} Design}
\label{sec:algo}

In this section, we model the learning problem and introduce \namex{}, our proposed learning framework based on RL. 

\subsection{Problem Modeling}

We assume a discrete-time system where time is divided into small steps. 
A \gls{UE} can change its state at every step. 
For the ease of exposition we first assume that an RL agent makes an observation of the environment and takes an action at every step. 
Later, we will show how the agent can be generalized to other cases with different temporal dynamics.

\textbf{Observation space.}
The RL agent operates under partial observability, as it omits variables like per-\gls{UE} mobility, \texttt{IDLE}/\texttt{ACTIVE} timers, and reselection counters.
Formally, the control task is modeled as a \gls{POMDP}.
The observation space for the agent consists of per-cell features and global metrics.
Per-cell features include the normalized available bandwidth and the number of connected \glspl{UE} in \texttt{ACTIVE} mode normalized by the total number of \glspl{UE}, including the ones camping on it in \texttt{IDLE} mode.
Global metrics include the mean and standard deviation of available bandwidth across all cells, the mean and standard deviation of the normalized number of connected \glspl{UE} per-cell, and the ratio of \glspl{UE} in \texttt{IDLE} mode to the total number of \glspl{UE}.
These global metrics compress the per-cell data into a fixed-size representation, providing a bird's-eye view of the network state.
This fixed-size representation allows the policy to generalize across varying network topologies, as it decouples global network state representation from the size.

\begin{table}[!t]
\caption{Cell (Re)Selection Parameters, Ranges, and Purposes}
\label{tab:reselection-params}
\centering
\begin{tabular}{@{}p{0.1\columnwidth} p{0.16\columnwidth} p{0.64\columnwidth}@{}}
\toprule
\textbf{Symbol} & \textbf{Range (dB/dBm)} & \textbf{Purpose} \\
\midrule
$T^{C_i \rightarrow C_j}_{\mathrm{XHigh}}$ & $[-100, 0]$ & Thresh. to reselect to higher priority cell \\
$T^{C_i \rightarrow C_j}_{\mathrm{XLow}}$ & $[-100, 0]$ & Thresh. to reselect to lower priority cell \\
$T^{C_i}_{\mathrm{SLow}}$ & $[-100, 0]$ & Serving cell signal strength reselection threshold  \\
$Q^{C_i}_{\mathrm{Hyst}}$ & $[0,\ 30]$ & Hysteresis to stabilize serving cell selection \\
$Q^{C_i \rightarrow C_j}_{\mathrm{Offset_{freq}}}$ & $[0,\ 30]$ & Offset biasing reselection btw. equal priority cells \\
$Q^{C_i}_{\mathrm{rxlevmin}}$ & $[-100, 0]$ & Min. required signal level for cell suitability \\
\bottomrule
\end{tabular}
\end{table}

\textbf{Action space.}
For each of the cell (re)selection parameters (see Table~\ref{tab:reselection-params}), the model outputs the mean $\mu_i$ and standard deviation $\sigma_i$ that are used to define a Gaussian distribution.
The action is sampled from these Gaussian distributions, allowing the agent to adaptively adjust the (re)selection parameters.
The action space is a continuous vector of size $2 \cdot 6$.
The constant six comes from the six (re)selection parameters that are shared across all cells.%
The first half of the action vector contains the means, and the second half contains the standard deviations for all the (re)selection parameters.
As the action controls both the mean and standard deviation, it allows the agent to adaptively adjust the exploration intensity by modifying the standard deviation.
The resulting actions (in range $[0,1]$) are then linearly mapped to the physical ranges for the parameters (see Table~\ref{tab:reselection-params}).
We intentionally exclude cell priorities from the action space because they introduce discrete, high-impact changes in the reselection logic and are typically adjusted on much longer time scales.
Cell priorities are fixed to the values used by the conventional heuristic baseline.

\textbf{Neural network architecture.}
To keep the agent fast and lightweight, we adopt a simple four-layer feedforward neural network. 
The input layer is designed to match the dimensions of the environment's observation space.
Following that, the network features a \gls{MLP} with two hidden layers and $tanh$ activation functions.
The output layer uses a sigmoid activation function to ensure that the actions remain within a valid range for the action space.

\textbf{Reward.}
The reward function is designed to encourage the agent to improve on the following three specified objectives: 
The \emph{throughput} term is used to reward the agent for achieving higher total throughput calculated on the basis of the total throughput of all \glspl{UE} in the network.
The \emph{load-balancing} term encourages the agent to equalize throughputs across individual cells.
The \emph{per-\gls{UE} efficiency} term is designed to encourage the agent to maximize the throughput of each \gls{UE} while considering the number of active \glspl{UE} in the network.
The agent receives higher reward, if per-\gls{UE} throughput is high while more \glspl{UE} are active, compared to where only few \glspl{UE} are active, to avoid the agent focusing on a few \glspl{UE} at the expense of others.

During training, the agent receives a reward at each step based on its action and the environment's response. 
The reward is a composite function designed to balance throughput, load balancing, and per-UE efficiency.
At each time step $t$, the reward is computed as
\begin{equation}
    r_t = w_1 \cdot r^{\text{tput}}_t + w_2 \cdot r^{\text{bal}}_t + w_3 \cdot r^{\text{ue-eff}}_t
\end{equation}
where $w_1 + w_2 + w_3 = 1$ are weighting coefficients.
The throughput term is defined by
\begin{equation}
    r^{\text{tput}}_t = \left( \frac{T_t}{B^{\text{tput}}(t)} \right) - 1
\end{equation}
where $T_{t}$ is the total throughput, and $B^{\text{tput}}(t)$ is a moving average baseline estimator.
The load-balancing term can be characterized by 
\begin{equation}
    r^{\text{bal}}_t = \left( \frac{B^{\sigma}(t)}{\sigma_t} \right) - 1
\end{equation}
where $\sigma_t$ is the standard deviation of per-cell throughputs, and $B^{\sigma}(t)$ is the corresponding baseline estimator. This is in line with the goal to provide the same expected throughput in a given area that operators aim to achieve.
Finally, the per-\gls{UE} efficiency term is given by 
\begin{equation}
    r^{\text{ue-eff}}_t = \left( \frac{\overline{T}_{\text{UE},t}}{B^{\text{UE}}(t)} - 1 \right) \cdot \min\left(1, \frac{\overline{\text{UE}}_{\text{active},t}}{\text{UE}_{\max}} \right)
\end{equation}
where $\overline{T}_{\text{UE},t}$ is the mean per-UE throughput, $B^{\text{UE}}(t)$ is its baseline estimator, and $\overline{\text{UE}}_{\text{active},t}$ is the average number of active \glspl{UE} during interval $t$ and $\text{UE}_{\max}$ is the maximum number of \glspl{UE} in the network.
This formulation rewards the agent for load scenarios where many \glspl{UE} are active, representing high-load network conditions.

\textbf{Baseline estimators.}
To determine whether the agent is performing better or worse, we employ baseline estimators, computed separately for each of the reward terms, as detailed above in the reward function definition.
The baseline estimators are used to compute the relative performance difference between the current agent performance and the baseline performance from the past. 
The calculation of the baseline estimators are explained in the following section. 
At initialization, when the baseline has not yet been populated, we use the performance of a conventional heuristic approach as a point of reference. 

\subsection{\namex{} Learning Algorithm and Optimizations}

The agent is trained on episodes each of which is defined as a set of consecutive steps in a specific environment. 
To ensure reproducibility, we seed the environment with a fixed random seed for each episode.
The seeds are used for random number generator initialization and determine the stochastic processes which control the \glspl{UE}' modes (\texttt{IDLE}/\texttt{ACTIVE}) and positions, effectively generating different network topologies and traffic patterns during training.
The environment seed is used to generate a unique scenario, including the cell topology and \glspl{UE} positions.
Each \gls{UE} receives a unique random number generator seed derived from the episode seed, based on which \gls{UE} seeds the location and mode throughout the episode.
\glspl{UE} are placed in realistic positions on streets or inside buildings and cell tower positions correspond to real cell tower positions.
For the mode switch, each \gls{UE} generates state-transition events drawn from exponential distributions seeded with the \gls{UE}-specific seed.
For evaluation, we use seeds that are unknown to the agent during training, presenting unique scenarios to the agent.
After each episode, we reset the environment.

During the development of the learning algorithm, we first experimented with the popular actor-critic method A2C~\cite{DBLP:conf/icml/MnihBMGLHSK16}, but it did not converge, likely due to the intrinsic dynamics of the problem.
We hypothesize that instability arises from the fact that different seeds yield different total throughput.
Each seed produces different temporal reward profiles, leading to non-stationary targets for the value function.
Hence, the critic network tries to regress to a mixture of unrelated target values.
Without a normalizing baseline, the advantage estimates vary significantly, leading to gradient interference.
A similar problem is described in~\cite{2019-sigcomm-decima}, where the limitations of value-based methods in scheduling contexts are mentioned.
Specifically, it is highlighted that if the target of the value function becomes non-stationary, the learning can be hindered.
To avoid this problem, we adopt the vanilla \texttt{REINFORCE} algorithm~\cite{1992-ml-gradient} with seed-specific baseline estimators~\cite{2018-arxiv-variancereduction}.
The \texttt{REINFORCE} algorithm does not use a shared critic network but computes the gradients based on the rewards received at each time step.
Since rewards are normalized by the baseline estimators, gradient interference between seeds is avoided.

The baseline estimators in the reward terms are used to capture the expected performance of the agent.
For each seed $s$, a baseline $B_s(t)$ is computed as a moving average over the last $w$ episodes at time step $t$.
The episodes with the same seed share the same network topology and traffic patterns, making the baseline an unbiased estimator of the expected return for the current policy.

To stabilize training, we apply global gradient clipping using the $\ell_2$ norm.
This prevents large updates that could destabilize the learning process~\cite{2023-icpram-gradientclipping}.
To favor known regions of the action space, we initialize the model weights of the neural network so that the initial mean actions are likely to reproduce the parameter set of the traditional approach.
We accomplish this by writing the values into the output-layer biases of the neural network.
This warm start helps the agent to quickly adapt to the environment and improve initial performance~\cite{2021-aaai-warmstart}.
During training, naturally, the agent's performance fluctuates.
To ensure that we retain the best-performing model, we implement a checkpointing mechanism that periodically saves the current model.
We determine convergence during training from the episode rewards.
If the reward approach zero, i.e., the agent cannot improve on top of history performance, we conclude training convergence.

\textbf{Curriculum learning.}
Initially, the agent may take poor actions, leading to sub-optimal resource allocation, not only for \glspl{UE} that switch to \texttt{ACTIVE} mode right after, but also for \glspl{UE} that become active later as the resources are blocked until released by the \glspl{UE}.
In such cases, the agent may not be able to correct its previous errors.
Even worse, the agent has to take sub-optimal actions based on the sub-optimal state, as such earlier errors may lead the policy into regions of the state space from which recovery is difficult or impossible.
To mitigate this, we employ a powerful technique called curriculum learning~\cite{2009-icml-curriculum}, starting with shorter episodes which provide a simpler environment.
The agent learns to take good actions from such shorter episodes, stabilizing early learning and avoiding sub-optimal states.
We then gradually increase the complexity of the environment by extending the episode length, allowing the agent to learn to take good actions from longer episodes.
During the training, the learning rate is set to $10^{-4}$ at the beginning of training and is adjusted by halving it with every new curriculum learning round.

\textbf{History stacking.}
As the problem is modeled as a \gls{POMDP}, we augment the observation space with a history of the last $k$ full observations 
to improve the state inference.
This history stacking allows the agent to disambiguate otherwise identical state observations, such as the same cell-load vector arising from rising versus falling load trends.
This reduces partial observability, improving decision making as long-term trends can be inferred from the history~\cite{2015-aaai-pomdp,2013-arxiv-atari,2024-ijcnn-deepstacking}.

%% file: text/generalization.tex
\section{Model Generalization}
\label{sec:generalization}

Generalizability is essential for the application of RL-based solutions in real world. \namex{} incorporates the following aspects in its design to address generalizability challenges. 

\textbf{Observation space design.}
The observation space features are expressed as ratios, e.g., per-cell available-to-total bandwidth.
Each value is normalized and hence is independent of the topology scale and absolute network load.
Additionally, we include global metrics: mean and standard deviation of available bandwidth, of connected \glspl{UE}, and the idle ratio (the ratio of \glspl{UE} in the \texttt{IDLE} mode to the total number of \glspl{UE}).
These design choices ensure that the observation distributions are scale-invariant, and the length of the observations vector grows linearly with the number of cells as we include a constant number of per-cell metrics.
Moreover, it allows the policy to generalize and transfer effectively to different regions with the same number of cell.

\textbf{Model design.}
By keeping a flat input vector, we can use a very small MLP (e.g., $2 \times 128$) instead of more complex GRU/LSTM architectures with significantly more parameters.
This choice leads to lower variance and markedly less overfitting to rare seeds, as the model has fewer parameters to adjust (cf. ``bias-variance tradeoff''~\cite{Vijayakumar2007BiasVariance,hastie2009elements}).
We use the AdamW~\cite{2019-iclr-regularization} optimizer, which is known to improve generalization by treating weight decay separately from adaptive learning rates for better regularization~\cite{2024-tpami-adamw}.

\textbf{Cross-scenario generalization.}
We employ domain randomization by generating \SI{100}{} unique seeds and scenarios with different \gls{UE} positions and traffic patterns.
The randomization of \glspl{UE} positions and traffic patterns leads to new load patterns in every episode, further enhancing the agent's ability to generalize.
This approach ensures that the agent is exposed to a wide range of scenarios during training, which helps the policy to achieve cross-traffic generalization.
The reward function incorporates seed-specific baseline estimators, which eliminate per-scenario bias.
This ensures that the gradients have similar magnitudes across different cell topologies or traffic intensities, aiding cross-seed transfer, as the agent learns to focus on relative improvements rather than absolute values
The agent is trained on a specific load, such as 500 \glspl{UE} where the state space is designed to be independent of the number of \glspl{UE}.
By design, the agent should be able to generalize to different loads, such as 200 or 700 \glspl{UE}.

\textbf{Episode-related generalization.}
The agent is trained with curriculum learning, starting with shorter episodes and gradually increasing the episode length.
This approach allows the agent to master short tasks first, stabilizing learning.
Once the agent has learned to take good actions from shorter episodes, it is exposed to longer episodes, which teaches the agent to plan over slower phenomena.
As a result, the agent should be able to cope with different episode lengths and control periods during inference.
To prevent the agent from memorizing the order of seeds, we shuffle the seed order during training.
This approach helps the agent generalize better and converge earlier, as it is exposed to a diverse set of scenarios in a non-sequential manner.
The shuffling of seeds ensures that the agent does not overfit to specific scenario sequences, allowing for improved generalization across different training episodes.

\textbf{Generalization over temporal dynamics.}
There are two important time scales in the system:
(1) the control frequency of the agent, which is the period between two reconfigurations of cell parameters, referred to as \gls{PRI},
(2) the \gls{UE} state dynamics, which is the average number of steps a \gls{UE} remains in a given mode (either \texttt{ACTIVE} or \texttt{IDLE}), referred to as \gls{SDT}.
There are three different cases for the ratio between \gls{PRI} and \gls{SDT}, which refer to as \gls{UDR}: 
\gls{UDR} $> 1$ where the agent reconfigures more frequently than \glspl{UE} change their mode, enabling rapid adaptation, 
\gls{UDR} $\approx 1$ where the agent and \glspl{UE} operate on comparable time scales, resulting in synchronized dynamics, and
\gls{UDR} $< 1$ where the agent acts on a slower time scale than the \glspl{UE}, potentially limiting responsiveness.
\namex{} is designed to infer long-term trends in network load and \gls{UE} behavior which mitigates partial observability by introducing a history stack in the observation space enabling the agent to distinguish between rising and falling load patterns.
We will show that \namex{} shows excellent capabilities in handling different \glspl{UDR}.

%% file: text/evaluation.tex
\section{Evaluation}
\label{sec:eval}

In this section, we evaluate the performance of \namex{} with real-world mobile network scenarios and aim to answer: (1) What are the performance gains of \namex{} compared to the conventional heuristic approach? 
(2) How well does \namex{} generalize to new network scenarios different from the ones used for training?
(3) How do the major optimizations applied in \namex{} contribute to the performance gains?

\subsection{Evaluation Setup}

We build a simulator in Python implementing the cellular network model described in Section~\ref{sec:problem-modeling}.
The simulator uses line-of-sight geometry and a sectorized antenna pattern with a \SI{120}{\degree}-sector width -- beamwidth was inferred from the spatial layout and visual coverage of cells, while azimuths and positions were provided by the network operator -- and a \SI{10}{\deci\bel} gain for the main lobe~\cite{BechtaImpact2020}, with a \SI{0}{\deci\bel} signal elsewhere as a simplifying modelling choice.
The free-space path-loss model is used at GHz frequencies, and we optionally include a light building obstruction loss of \SI{6}{\deci\bel}~\cite{miao2025_midband_fr3_6g_survey} for each wall that intersects the line-of-sight between a cell and a \gls{UE}.
Thermal noise is modeled as \SI{-174}{\deci\bel\milli\watt\per\hertz} (Johnson-Nyquist noise~\cite{rosu2023_understanding_noise_figure}), and the spectral efficiency is computed using a nearest-neighbor lookup based on the \glsentryshort{SINR}-to-\glsentryshort{SE} mapping~\cite{2024-access-stateupdate}, assuming a noise-limited environment.
We use a resource-fair scheduling algorithm that allocates bandwidth equally to all active \glspl{UE} in a cell, iteratively ensuring fairness using a water-filling algorithm.

\begin{figure}[t]
  \centering
  \includegraphics[width=0.9\columnwidth]{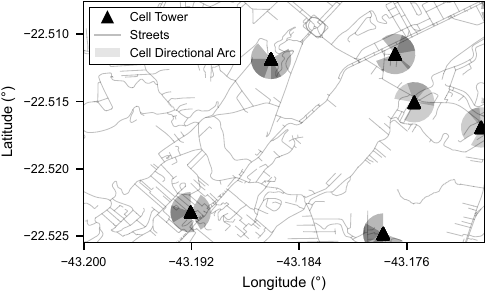}
  \caption{Large network deployment area used for training and evaluation~\cite{osm2025}.}
  \label{fig:map-large}
\end{figure}

Our experiments are conducted over multiple distinct geographical areas, each with a unique topology.
The cell tower information is extracted from a real-world dataset obtained from a major mobile operator in Brazil. 
The dataset provides exact locations, and deployment configurations such as azimuth, beamwidth, frequency, bandwidth, and reselection parameters.
Since not all reselection parameters are available in the dataset, we infer representative values for missing parameters sourced from publicly available material~\cite{techplayon2024cellreselection,sirdc_lte_parameter_engineering_2025,prasadnandkumar2025lte_ran_parameter_mapping} that are not specified for the conventional approach and the warm start of \namex{}.
Industry guideline~\cite{ngmn2008_son_oam_requirements} validates our approach of deriving default parameters from multiple authoritative sources, including vendor specifications and industry best practices, when complete operator configuration datasets are unavailable.
We use the following three geographical areas located in Petr\'{o}polis, Brazil in our experiments.
\begin{itemize}
    \item \textbf{Baseline area} consists of two cell towers, 15 cells, and is of size \SI{2.36}{\kilo\metre\squared}.
    \item \textbf{Alternative area} is used for evaluation only and has the same number of cell towers, number of cells, and area size as the baseline area, but is geographically distinct.
    \item \textbf{Large area} is significantly larger with six cell towers (see Figure~\ref{fig:map-large}), 48 cells, and is stretched across \SI{6.10}{\kilo\metre\squared}. This area is used to evaluate \namex{} scalability.
\end{itemize}

\textbf{Temporal dynamics.}
We evaluate the RL agent under different temporal dynamics to assess its robustness and adaptability.
The baseline model is trained with a \gls{PRI} of one and evaluated with the same \gls{PRI}, resulting in a reference \gls{UDR} of approximately \SI{0.2}{}.
The \textit{Stress Test} scenario evaluates the baseline model with a \gls{PRI} of \SI{10}{} and a \gls{UDR} of approximately \SI{2}{}.
This scenario tests the robustness of the RL agent against previously unseen temporal dynamics, where on average two \gls{UE} mode changes can occur between two updates.
The \textit{Synchronous Updates} scenario trains the model with a \gls{PRI} of \SI{5}{} to match the empirical \gls{SDT}, resulting in a \gls{UDR} of approximately one.
In the \textit{Slow Updates} scenario, the model is trained and evaluated with a \gls{PRI} of \SI{10}{} to consistently simulate a slow control loop, resulting in a \gls{UDR} of approximately \SI{2}{}.

\textbf{UE setup.}
The \glspl{UE} are placed on valid positions on streets or inside buildings in the simulated area, ensuring realistic positions.
The simulalator also supports building obstruction loss, which can be enabled to simulate more realistic scenarios.
For performance reasons, we disable building obstruction loss during training to accelerate simulation and training to allow for more extensive exploration of diverse scenarios.
The \glspl{UE}' mode switch is modeled as a Poisson process with exponential interarrival times parameterized with rates $\lambda_\text{IDLE} = \lambda_\text{ACTIVE}= 0.2$ leading to an average \gls{SDT} of \SI{5}{\second}.
UE mobility is supported by our simulator but disabled in our simulations because \namex{} does not consider handovers in the current design.
However, the agent is trained on a variety of different scenarios with unique \gls{UE} positions.
The training set consists of \SI{100}{} unique scenarios, each with \SI{500}{} \glspl{UE}.
The \gls{UE} positions and traffic patterns are seeded randomly, ensuring that the agent is exposed to a diverse set of conditions.
The validation is performed against \SI{100}{} unseen scenarios, which allows us to evaluate the generalizability of the agent across different \gls{UE} distributions and traffic patterns.

\textbf{Model and system parameters.}
The models use \SI{1024}{} neurons per layer with \SI{2}{} hidden layers.
During our training experiments, we observed that \SI{512}{} neurons per layer led to unstable training dynamics and suboptimal performance.
To ensure stability during training, we apply gradient clipping with a threshold of \SI{10}{}.
The action space ranges are normalized to the values depicted in Table~\ref{tab:reselection-params}.
The weights of the reward components are assigned as follows: $w_1 = 0.4$ for throughput, $w_2 = 0.4$ for load balancing, and $w_3 = 0.2$ for per-UE efficiency.
For the two scenarios--the scalability evaluation on the large area (Figure~\ref{fig:eval-large}) and the synchronous updates case in the temporal dynamics evaluation (Figure~\ref{fig:eval-time})--we employ reward weights $w_1 = 0.025$, $w_2 = 0.95$, and $w_3 = 0.025$.
In these scenarios, the agent achieved proportionally larger throughput gains more easily; hence, we compensate with a higher load-balancing weight to encourage more balanced performance improvements.
To limit exploration intensity, we use a Gaussian distribution with a standard deviation capped at \SI{0.1}{} for the action space.
For curriculum learning the initial episode length is set to \SI{30}{\second}, and is incremented by \SI{10}{\second} after \SI{3}{} passes through all environment seeds.
The increase is applied \SI{4}{} times in total leading to a final episode length of \SI{50}{\second}.
History stacking is applied, where the last $k = \SI{10}{}$ full observations are included in the observation vector.

\subsection{Overall Performance \& Generalizability}

We first evaluate \namex{} performance
(1) across the three geographical areas 
(2) under load scaling from \SI{100}{} to \SI{1000}{} \glspl{UE}, and 
(3) over four different temporal dynamics from low-latency to slow control loops (\textit{Baseline}, \textit{Stress Test}, \textit{Synchronous Updates}, and \textit{Slow Updates}),
thus probing spatial, load-dependent and temporal generalization capabilities of our approach.
For each evaluation we confront the agent with \SI{100}{} unique scenarios which are unseen during training with \SI{500}{} \glspl{UE} with a fixed episode length of \SI{60}{} seconds.
To investigate load scaling, we evaluate the agent on the baseline area with \SI{100}{} to \SI{1000}{} \glspl{UE} in increments of \SI{100}{}.
We use the conventional approach as a baseline for comparison using the reference values in Table~\ref{tab:heuristic-reference}.
For easier visual interpretation, the load-balancing component in the reward term---defined as a (normalized) standard deviation where lower is better---is computed as its reciprocal $1/r^{\text{bal}}_t$, so that positive values denote improvement.

\begin{figure}[!t]
  \centering
  \includegraphics[width=0.9\columnwidth]{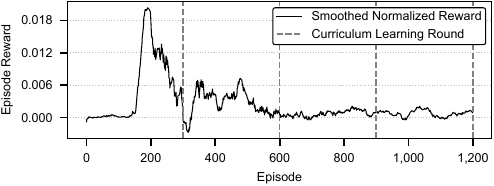}
  \caption{Achieved total episode reward over episodes using the baseline area.}
  \label{fig:eval-convergence}
\end{figure}

\textbf{Convergence.}
Due to the moving-average estimators, the same performance yields a lower reward as training progresses, which requires the agent to improve to maintain the same reward levels.
Naturally, the agent's reward decreases over the training process.
When the reward approaches zero, the agent converges.
Sometimes, this is ambiguous as phases at zero reward are followed by further learning.
To decide on convergence, we check throughput and load-balancing performance for models at different stages.
We observed that sometimes the agent gets stuck at zero reward for a while but then learns further, optimizing only one metric.
Throughput gains are relatively easier to achieve and larger; hence, we adapt reward weights to compensate in certain scenarios as described above.
Figure~\ref{fig:eval-convergence} shows the episode reward normalized by the episode length.
Vertical lines indicate the curriculum learning rounds, where the episode length is increased.
The reward is smoothed by an exponentially weighted moving average ($\alpha=0.2$) over the training.
During the first round of curriculum learning, the normalized episode rewards show fluctuations between \SI{0.02}{} and \SI{-0.004}{}.
Training stabilizes after the second round of curriculum learning, indicating convergence.
The standard deviation over the last \SI{100}{} episodes is \SI{7e-4}{}.

\begin{figure}[!t]
  \centering
  \includegraphics[width=0.9\columnwidth]{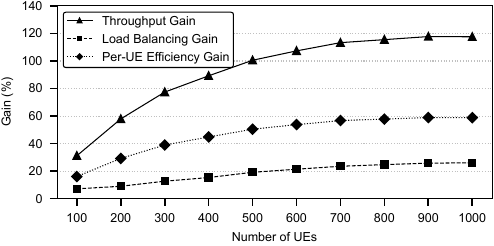}
  \caption{Average values for reward components for the baseline area over different numbers of \glspl{UE} in the system. The agent is trained on \SI{500}{} \glspl{UE}.}
  \label{fig:eval-baseline}
\end{figure}

\textbf{\gls{UE} scaling.}
We quantify how performance varies with network load by fixing the agent trained on \SI{500}{} \glspl{UE} and evaluating it on $N\!=\!100,200,\dots,1000$ \glspl{UE} in the baseline area.
Figure~\ref{fig:eval-baseline} shows the components of the reward function separately as average values over the number of \glspl{UE} in the system.
Despite being trained on \SI{500}{} \glspl{UE}, the agent can generalize to different numbers of \glspl{UE}, showing a gain for all components of the reward function.
It consistently shows better average throughput, load balancing, and per-\gls{UE} efficiency compared to the conventional heuristic across all evaluated numbers of \glspl{UE}.
In general, the gains are lower for smaller $N$.
Load-balancing improves up to \SI{26}{\percent} for $N=1000$ compared to the conventional heuristic.
Average total throughput consistently improves with increasing number of \glspl{UE} with a maximum gain of \SI{118}{\percent}.
Average per-\gls{UE} efficiency improves up to \SI{59}{\percent} for $N=1000$ compared to the conventional heuristic.
In general, higher numbers of \glspl{UE} lead to diminishing returns for all three components.
For throughput and per-UE efficiency, this is expected since the conventional heuristic increasingly saturates the available spectrum as \gls{UE} count grows.
Similarly, load balancing naturally improves with more \glspl{UE} per cell, reducing the optimization potential for the agent.

\begin{figure}[!t]
  \centering
  \includegraphics[width=0.9\columnwidth]{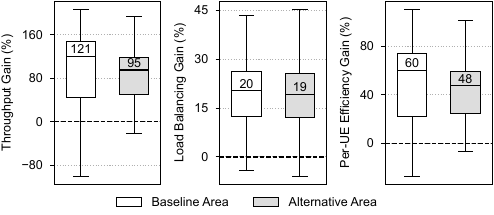}
  \caption{Performance gains for the baseline area and alternative area. The agent is trained on the baseline area and evaluated on both areas to show the cross-area generalizability.}
  \label{fig:eval-base-alt}
\end{figure}

\textbf{Cross-area evaluation.}
We use the agent trained on the baseline area and evaluate it on the same baseline area as well as the alternative area, which has a different cell topology and \gls{UE} positions but share geographic area size, number of cell towers and cells.
Figure~\ref{fig:eval-base-alt} depicts the results of the cross-area evaluation.
The agent outperforms the conventional heuristic in both areas across all three metrics.
The performance gains in the alternative area are comparable to those achieved in the baseline area.
The confidence intervals for throughput and per-\gls{UE} efficiency are smaller in the alternative area, indicating more consistent performance than in the baseline area.
This indicates that the agent has learned a policy that generalizes well to unseen areas with different cell topologies and \gls{UE} distributions.
We attribute this to the geographical features of the alternative area, where cells are more evenly distributed over the area compared to the baseline area, where cells are more clustered in certain parts of the area leading to more diverse \gls{UE}-to-cell associations.
In total, this shows the robustness of the policy to topology shifts.

\begin{figure}[t]
  \centering
  \includegraphics[width=0.9\columnwidth]{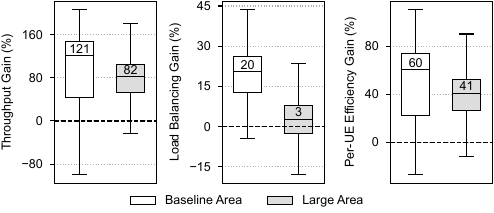}
  \caption{Performance gains for the large area. The agent is trained on the large area and evaluated on it. The baseline-area model is shown for comparison to highlight the scalability of the approach.}
  \label{fig:eval-large}
\end{figure}

\textbf{Scalability to larger areas.}
To evaluate the scalability of policy, we train the agent on the large area with about $2.6 \times$ the size of the baseline area and $3.2 \times$ the number of cells and $N = 500$ \glspl{UE}.
For this scenario, we increase the moving-average estimator window size from \SI{2}{} to \SI{10}{} to provide a more stable baseline for reward computation.
We observed that with the smaller window size, the baseline prediction adapts too quickly to bad performance, resulting in rewards that remain close to zero and provide insufficient learning signal.
In contrast, a larger window size maintains a higher baseline estimate that does not adapt as quickly, requiring the agent to achieve larger performance improvements to generate positive rewards and sustain learning progress.
Figure~\ref{fig:eval-large} shows the results of the evaluation on the large area.
The agent outperforms the conventional heuristic across all metrics with high confidence.
The median value for total throughput gain is smaller at \SI{82}{\percent} compared to \SI{121}{\percent} in the baseline area.
This is expected as with fewer \glspl{UE} per cell the conventional heuristic already uses most available spectrum, hence, the agent has limited headroom to increase total throughput.
The load-balancing component shows the modest improvement with a median value of \SI{3}{\percent} over the conventional heuristic, about $1/6$ of the gain in the baseline area.
This is expected as we use homogeneous (re)selection parameters across all cells, which limits the agent's ability to fine-tune load balancing in a larger area with more cells.
The per-\gls{UE} efficiency gain shows a median value of \SI{41}{\percent} over the conventional heuristic which is around two thirds of the gain in the baseline area.
Even in larger network topologies, our approach reliably improves over the conventional heuristic proving its scalability.

\begin{figure}[t]
  \centering
  \includegraphics[width=0.9\columnwidth]{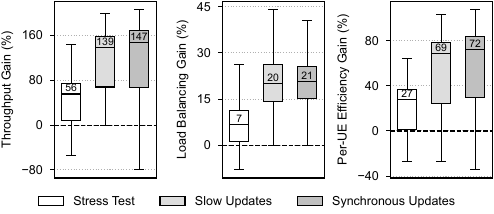}
  \caption{Performance gains for different temporal dynamics. In the \textit{Stress Test} scenario, the agent is trained with a $\text{UDR} > 1$ and evaluated with a $\text{UDR} < 1$. The \textit{Slow Updates} scenario trains and evaluates the agent with a $\text{UDR} < 1$. The \textit{Synchronous Updates} scenario trains the agent with a $\text{UDR} \approx 1$.}
  \label{fig:eval-time}
\end{figure}

\textbf{Diverse temporal dynamics.}
Figure~\ref{fig:eval-time} shows the results under different temporal dynamics.
The \textit{Stress Test} scenario shows a median of \SI{56}{\percent} in throughput gain, \SI{7}{\percent} in load balancing, and \SI{27}{\percent} in per-\gls{UE} efficiency.
The gains are lower than in the baseline scenario where the agent is evaluated under the same temporal dynamics as it was trained on.
Intuitively this makes sense, as the agent learned to take action on shorter timescales and is now confronted with a slower control loop.
The agent cannot react as quickly to traffic bursts and update network parameters.
Compared to the conventional heuristic, however, the agent still manages to improve performance across all three metrics, indicating that it has learned a policy that is robust to changes in temporal dynamics.
The \textit{Slow Updates} scenario shows a median throughput gain of \SI{139}{\percent}, \SI{20}{\percent} load balancing gain and \SI{69}{\percent} for per-\gls{UE} efficiency yielding slightly higher gains in throughput and per-\gls{UE} efficiency compared to the baseline scenario.
These results suggest that by training on a slower timescale, the agent learns to identify more resilient, long-term parameter configurations that generalize better across multiple user state transitions.
The \textit{Synchronous Updates} scenario shows a median gain of \SI{147}{\percent} in total throughput, \SI{21}{\percent} in load balancing, and \SI{72}{\percent} in per-\gls{UE} efficiency.
These results outperform all other temporal dynamics scenarios.
This indicates that aligning the update interval with the timescale of \gls{UE} transitions allows the agent to optimally time parameter updates, minimizing reactive lag and enhancing overall network performance.

\subsection{Ablation Study}
\label{sec:ablation}

\begin{figure}[t]
    \centering
    \includegraphics[width=0.9\columnwidth]{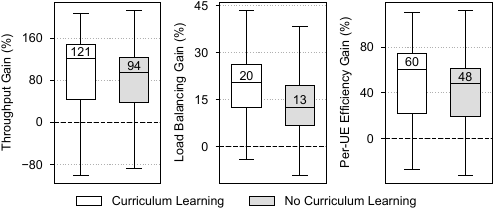}
    \caption{Performance gains for the baseline area without curriculum learning. The agent is trained with a fixed episode length of \SI{50}{\second}.}
    \label{fig:no-curriculum}
\end{figure}

\textbf{Curriculum learning.}
We evaluate the impact of curriculum learning by disabling it and setting the episode length to \SI{50}{\second} for all episodes, i.e., the final episode length with curriculum learning.
We also keep the repeating mechanism, hence, the agent is trained on each unique episode three times in a random order.
Figure~\ref{fig:no-curriculum} shows the results of this evaluation with performance gains across all metrics.
The standard deviation of the episode reward over the last \SI{100}{} episodes is 4.9$\times$ higher compared to the curriculum learning enabled scenario, indicating that the policy converges slower.

\begin{figure}[t]
    \centering
    \includegraphics[width=0.9\columnwidth]{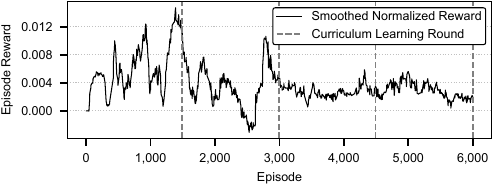}
    \caption{Total episode reward over episodes  (baseline area, \SI{500}{} seeds).}
    \label{fig:convergence-500-seeds}
\end{figure}

\begin{figure}[!t]
    \centering
    \includegraphics[width=0.9\columnwidth]{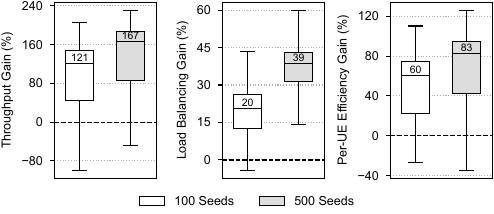}
    \caption{Performance gains for the baseline area with \SI{500}{} seeds. The \SI{100}{}-seed model is shown for comparison to highlight the impact of increased training diversity.}
    \label{fig:eval-500-seeds}
\end{figure}

\textbf{Number of episode seeds.}
We evaluate the impact of the number of episode seeds used for training.
The default number of seeds is \SI{100}{}.
We increase this number to \SI{500}{}.
When using \SI{500}{} seeds, the agent converges noticeably slower due to the larger diversity of scenarios.
Figure~\ref{fig:convergence-500-seeds} shows the episode rewards.
Even during the last two curriculum learning rounds learning progress is still visible.
We evaluate the model at two checkpoints:
After \SI{2300}{} episodes, when the agent shows first signs of convergence and reward values are close to zero and at the end of training episodes to assess the remaining learning potential.
Unlike the \SI{100}{}-seed case, the late checkpoint substantially outperforms the early checkpoint in all three metrics, achieving balanced performance gains rather than optimizing selectively for one metric.
The early model trained on \SI{500}{} seeds demonstrates intermediate performance: It surpasses the \SI{100}{}-seed model but underperforms relative to the late 500-seed model.
The last checkpoint (Figure~\ref{fig:eval-500-seeds}) achieves median throughput gains of \SI{167}{\percent}, load-balancing gains of \SI{39}{\percent}, and per-\gls{UE} efficiency gains of \SI{83}{\percent}.
The performance improves across all three metrics, with load balancing achieving the largest relative gain, as expected when a more diverse training dataset is presented to the agent during training.

\textbf{UE Mobility.}
We evaluate the impact of \gls{UE} mobility by enabling it during evaluation only while keeping training on static \glspl{UE}.
\glspl{UE} move at an average speed of \SI{30}{\kilo\metre\per\hour} along streets--\glspl{UE} inside buildings do not move.
We evaluate two cases
(1) the baseline model is trained on \SI{100}{} and
(2) the model is trained on \SI{500}{} seeds to assess whether increased training diversity improves robustness to mobility.

\begin{figure}[t]
    \centering
    \includegraphics[width=0.9\columnwidth]{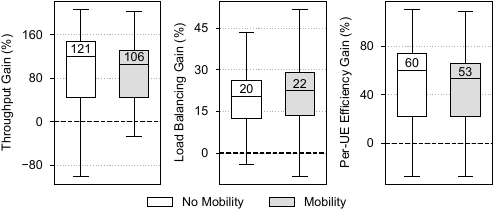}
    \caption{Performance gains for the baseline area with \gls{UE} mobility. The agent is trained on static \glspl{UE}. For comparison, the static \gls{UE} model is shown to highlight the impact of \gls{UE} mobility during evaluation.}
    \label{fig:eval-mobility}
\end{figure}

Figure~\ref{fig:eval-mobility} shows the results for the baseline model.
Even though the agent was trained exclusively on static \glspl{UE}, it still outperforms the conventional heuristic across all three metrics with a throughput gain of \SI{106}{\percent}, load-balancing gain of \SI{22}{\percent}, and per-\gls{UE} efficiency gain of \SI{53}{\percent}.
The consistency of these gains suggests that the agent has not merely memorized static \gls{UE} positions but has learned the underlying dynamics of the environment, enabling it to adapt effectively to \gls{UE} mobility.

\begin{figure}[t]
    \centering
    \includegraphics[width=0.9\columnwidth]{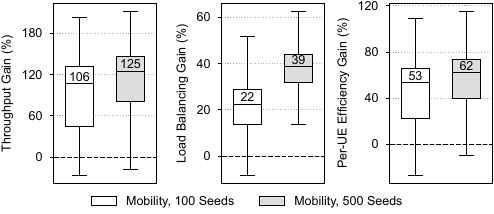}
    \caption{Performance gains for the baseline area with \gls{UE} mobility. The agent is trained on static \glspl{UE} with \SI{500}{} seeds. For comparison, \SI{100}{}-seed mobility model is shown to highlight the impact of increased training diversity under \gls{UE} mobility.}
    \label{fig:eval-500-seeds-mobility}
\end{figure}

Figure~\ref{fig:eval-500-seeds-mobility} shows the results for the model trained on \SI{500}{} seeds with \gls{UE} mobility during evaluation.
The performance further improves across all three metrics compared to the baseline model and the \SI{100}{}-seed model with mobility.
The throughput gain increases to \SI{125}{\percent}, load-balancing gain to \SI{39}{\percent}, and per-\gls{UE} efficiency gain to \SI{62}{\percent}.
The substantial improvements, particularly in load balancing, indicate that training on a more diverse set of scenarios can enhance the agent's ability to generalize to dynamic conditions such as \gls{UE} mobility.

%% file: text/discussion.tex
\section{Discussion}
\label{sec:discussion}

\textbf{Real-time parameter reconfiguration.}
Since \namex{} conducts cell (re)selection through parameter configurations, it is compatible with current infrastructure and does not require major physical update~\cite{2021-icnp-cellselection}. Yet, real-time parameter reconfiguration is still not straightforward with a large number of cells, given that \namex{} needs to collect the \gls{UE} state information from the network. Our evaluation already covers cases with less frequent reconfigurations. Nevertheless, deploying and testing \namex{} in a real network requires deep collaboration with mobile operators, which we are currently seeking.

\textbf{Heterogeneous cell parameters.}
\namex{} assumes homogeneous cell parameters across a network region (e.g., a district of a city), which is in line with how mobile networks are operated today. This simplification can largely be attributed to the high complexity involved in manual parameter configuration. \namex{} can be extended by expanding the state/action spaces to support heterogeneous parameters across different cells. 
A more advanced approach would be to introduce an RL agent for each cell and apply a multi-agent learning approach where each cell-local agent learns individually and shares knowledge by gossip-based state dissemination across cells.  

\textbf{Parameter-free cell (re)selection.}
\namex{} learns to configure parameters and does not select the cells directly. Staying with the parameters improves compatibility in real networks, but may limit the exploration space of the agent. Alternatively, we can train an agent for each \gls{UE} and use the local agent to make cell (re)selection decisions directly on each \gls{UE}, as studied in~\cite{2024-tnsm-multipoint} for multi-point coordination in mobile networks. Yet, the deployment of such solutions require infrastructure changes and may not be possible in the foreseeable future.

%% file: text/relatedwork.tex
\section{Related Work}
\label{sec:relatedwork}

\textbf{Cell (re)selection in mobile networks.} 
Cell (re)selection parameter optimization has received limited attention in the literature~\cite{2016-wcnc-layer,2020-mobicom-icellspeed,2021-icnp-cellselection}. 
To mitigate the performance issue observed in~\cite{2020-hotmobile-cellselection,2020-mobicom-icellspeed}, researchers have explored device-assisted fixes~\cite{2020-mobicom-icellspeed}. Some works leverage large datasets to learn how to configure newly added radio channels or tune parameters for better performance\cite{2021-sigcomm-auric, 2019-infocom-collaborative, 2022-imc-aurora, 2023-mobicom-chroma}. 
However, these techniques typically rely on static configurations and, in some cases, require manual verification. The closest work to ours is~\cite{2021-icnp-cellselection}, where they optimize the cell (re)selection parameters and propose a heuristic approach based on historical network loads. There are also studies focusing on handover optimizations~\cite{2018-imc-mobility,2016-sigmetrics-mobility,2020-sigcomm-5gmobility,2025-infocom-handover} or band switching~\cite{2024-hotmobile-bandswitching} for mobility management. To the best of our knowledge, we are the first to present a learning-based approach for adaptive parameter reconfiguration for cell (re)selection based on runtime network state. 

\textbf{RL for networking problems.} Many networking problems are intrinsically complex involving dynamic environments, which renders RL a powerful tool for tackling them~\cite{2019-csur-rlnet}. For example, researchers have applied RL-based solutions to problems such as video streaming~\cite{2017-sigcomm-pensieve}, congestion control~\cite{2024-infocom-rlcc}, and traffic scheduling~\cite{2024-infocom-fumes}. For mobile network management, RL has been applied to problems like network slicing~\cite{2021-pimrc-slicing} and handover optimization~\cite{2025-infocom-handover}. So far, no work has explored RL-based parameter reconfiguration for cell (re)selection.

%% file: text/conclusion.tex
\section{Conclusion}
\label{sec:conclusion}
In this paper, we presented \namex{}, a RL-based framework that autonomously adapts idle-mode cell (re)selection parameters in heterogeneous 4G/5G networks.
By modeling the problem as a \glsentrylong{POMDP} and incorporating optimization techniques 
\namex{} optimizes network performance under dynamic conditions.
Evaluation using a custom simulator with real-world topologies shows that \namex{} outperforms conventional heuristics by up to \SI{167}{\percent} in throughput and \SI{39}{\percent} in load balancing with robust generalizability across unseen geographic areas, scales, loads, 
and control loop latencies 
without retraining.
Compatible with existing 4G/5G specifications, \namex{} lays the groundwork for adaptive cell (re)selection, demonstrating the feasibility and benefits of transitioning from static configurations to RL-driven, autonomous network operations.

%% file: main.bbl
\begin{thebibliography}{10}
\providecommand{\url}[1]{#1}
\csname url@samestyle\endcsname
\providecommand{\newblock}{\relax}
\providecommand{\bibinfo}[2]{#2}
\providecommand{\BIBentrySTDinterwordspacing}{\spaceskip=0pt\relax}
\providecommand{\BIBentryALTinterwordstretchfactor}{4}
\providecommand{\BIBentryALTinterwordspacing}{\spaceskip=\fontdimen2\font plus
\BIBentryALTinterwordstretchfactor\fontdimen3\font minus
  \fontdimen4\font\relax}
\providecommand{\BIBforeignlanguage}[2]{{%
\expandafter\ifx\csname l@#1\endcsname\relax
\typeout{** WARNING: IEEEtran.bst: No hyphenation pattern has been}%
\typeout{** loaded for the language `#1'. Using the pattern for}%
\typeout{** the default language instead.}%
\else
\language=\csname l@#1\endcsname
\fi
#2}}
\providecommand{\BIBdecl}{\relax}
\BIBdecl

\bibitem{bna5GSA}
\BIBentryALTinterwordspacing
Bundesnetzagentur. (2024) Bundesnetzagentur extends mobile communications map
  to include high-performance 5g standalone. [Online]. Available:
  \url{https://www.bundesnetzagentur.de/SharedDocs/Pressemitteilungen/EN/2024/20240613_MoFu.html}
\BIBentrySTDinterwordspacing

\bibitem{fccSpectrum2022}
\BIBentryALTinterwordspacing
NTIA. (2022) Ntia establish spectrum coordination initiative. [Online].
  Available: \url{https://www.ntia.gov/press-release/2022/fcc-ntia-
  establish-spectrum-coordination-initiative}
\BIBentrySTDinterwordspacing

\bibitem{2024-hotmobile-bandswitching}
A.~Hassan, W.~Ye, A.~Zhang, J.~Carpenter, R.~Zhu, S.~Jin, F.~Qian, Z.~M. Mao,
  and Z.~Zhang, ``The case for boosting mobile application qoe via smart band
  switching in 5g/xg networks,'' in \emph{{ACM HotMobile}}, 2024.

\bibitem{2016-wcnc-layer}
M.~Malmirchegini, M.~Shukair, P.~Rached, S.~Sawhney, M.~Ambriss, K.~R.
  Chaudhuri, and S.~Sarkar, ``Layer management through idle-mode parameter
  optimization in multi-carrier {LTE} networks,'' in \emph{{IEEE} {WCNC}},
  2016.

\bibitem{2021-icnp-cellselection}
Q.~Li and C.~Peng, ``Reconfiguring cell selection in 4g/5g networks,'' in
  \emph{{IEEE ICNP}}, 2021.

\bibitem{2020-hotmobile-cellselection}
H.~Deng, K.~Ling, J.~Guo, and C.~Peng, ``Unveiling the missed 4.5g performance
  in the wild,'' in \emph{{ACM HotMobile}}, 2020.

\bibitem{3GPP2018-br}
{3GPP}, ``{5G}; {NR}; user equipment ({UE}) procedures in idle mode and in
  {RRC} inactive state,'' Oct. 2018.

\bibitem{3GPP2019-dn}
------, ``{LTE}; evolved universal terrestrial radio access ({E}-{UTRA}); user
  equipment ({UE}) procedures in idle mode,'' May 2019.

\bibitem{3GPP2017-rl}
------, ``Digital cellular telecommunications system (phase 2+) ({GSM});
  universal mobile telecommunications system ({UMTS}); {LTE}; vocabulary for
  {3GPP} specifications,'' Jul. 2017.

\bibitem{DBLP:conf/icml/MnihBMGLHSK16}
V.~Mnih, A.~P. Badia, M.~Mirza, A.~Graves, T.~P. Lillicrap, T.~Harley,
  D.~Silver, and K.~Kavukcuoglu, ``Asynchronous methods for deep reinforcement
  learning,'' in \emph{{ACM ICML}}, 2016.

\bibitem{2019-sigcomm-decima}
H.~Mao, M.~Schwarzkopf, S.~B. Venkatakrishnan, Z.~Meng, and M.~Alizadeh,
  ``Learning scheduling algorithms for data processing clusters,'' in
  \emph{{ACM} {SIGCOMM}}, 2019.

\bibitem{1992-ml-gradient}
R.~J. Williams, ``Simple statistical gradient-following algorithms for
  connectionist reinforcement learning,'' \emph{Mach. Learn.}, vol.~8, pp.
  229--256, 1992.

\bibitem{2018-arxiv-variancereduction}
H.~Mao, S.~B. Venkatakrishnan, M.~Schwarzkopf, and M.~Alizadeh, ``Variance
  reduction for reinforcement learning in input-driven environments,''
  \emph{CoRR}, vol. abs/1807.02264, 2018.

\bibitem{2023-icpram-gradientclipping}
A.~Ramaswamy, ``Gradient clipping in deep learning: {A} dynamical systems
  perspective,'' in \emph{{ICPRAM}}, 2023.

\bibitem{2021-aaai-warmstart}
A.~Silva and M.~C. Gombolay, ``Encoding human domain knowledge to warm start
  reinforcement learning,'' in \emph{{AAAI}}, 2021.

\bibitem{2009-icml-curriculum}
Y.~Bengio, J.~Louradour, R.~Collobert, and J.~Weston, ``Curriculum learning,''
  in \emph{{ACM ICML}}, vol. 382, 2009.

\bibitem{2015-aaai-pomdp}
M.~J. Hausknecht and P.~Stone, ``Deep recurrent q-learning for partially
  observable mdps,'' in \emph{{AAAI}}, 2015.

\bibitem{2013-arxiv-atari}
V.~Mnih, K.~Kavukcuoglu, D.~Silver, A.~Graves, I.~Antonoglou, D.~Wierstra, and
  M.~A. Riedmiller, ``Playing atari with deep reinforcement learning,''
  \emph{CoRR}, vol. abs/1312.5602, 2013.

\bibitem{2024-ijcnn-deepstacking}
K.~Jiang, Q.~Wang, Y.~Xu, and H.~Deng, ``Observation-time-action deep stacking
  strategy: Solving partial observability problems with visual input,'' in
  \emph{{IJCNN}}, 2024.

\bibitem{Vijayakumar2007BiasVariance}
\BIBentryALTinterwordspacing
S.~Vijayakumar, ``The bias--variance tradeoff,'' University of Edinburgh,
  Lecture Notes, 2007, original source no longer available; archived copy
  accessed 19~August~2014. [Online]. Available:
  \url{https://web.archive.org/web/20131005010514/http://www.inf.ed.ac.uk/teaching/courses/mlsc/Notes/Lecture4/BiasVariance.pdf}
\BIBentrySTDinterwordspacing

\bibitem{hastie2009elements}
\BIBentryALTinterwordspacing
T.~Hastie, R.~Tibshirani, and J.~Friedman, \emph{The Elements of Statistical
  Learning: Data Mining, Inference, and Prediction}, 2nd~ed., ser. Springer
  Series in Statistics.\hskip 1em plus 0.5em minus 0.4em\relax New York:
  Springer, 2009, accessed: 2025-07-30. [Online]. Available:
  \url{https://link.springer.com/book/10.1007/978-0-387-84858-7}
\BIBentrySTDinterwordspacing

\bibitem{2019-iclr-regularization}
I.~Loshchilov and F.~Hutter, ``Decoupled weight decay regularization,'' in
  \emph{{ICLR}}.\hskip 1em plus 0.5em minus 0.4em\relax OpenReview.net, 2019.

\bibitem{2024-tpami-adamw}
P.~Zhou, X.~Xie, Z.~Lin, and S.~Yan, ``Towards understanding convergence and
  generalization of adamw,'' \emph{{IEEE} Trans. Pattern Anal. Mach. Intell.},
  vol.~46, no.~9, pp. 6486--6493, 2024.

\bibitem{BechtaImpact2020}
K.~Bechta, C.~Grangeat, and J.~Du, ``Impact of effective antenna pattern on
  radio frequency exposure evaluation for 5g base station with directional
  antennas,'' in \emph{2020 XXXIIIrd General Assembly and Scientific Symposium
  of the International Union of Radio Science}, 2020.

\bibitem{miao2025_midband_fr3_6g_survey}
H.~Miao, J.~Zhang, P.~Tang, J.~Meng, Q.~Zhen, X.~Liu, E.~Liu, P.~Liu, L.~Tian,
  and G.~Liu, ``A survey of new mid‑band/fr3 for 6g: Channel measurement,
  characterization and modeling in outdoor environment,'' arXiv preprint
  arXiv:2504.06727 v1, 2025.

\bibitem{rosu2023_understanding_noise_figure}
I.~R. (VA3IUL), ``Understanding noise figure,''
  \url{https://www.qsl.net/va3iul/Noise/Understanding%20Noise%20Figure.pdf},
  2023, accessed: 2025-07-30.

\bibitem{2024-access-stateupdate}
E.~V. Markova, V.~E. Manaeva, E.~Zhbankova, D.~Moltchanov, P.~Balabanov,
  Y.~Koucheryavy, and Y.~Gaidamaka, ``Performance-utilization trade-offs for
  state update services in 5g {NR} systems,'' \emph{{IEEE} Access}, vol.~12,
  pp. 129\,789--129\,803, 2024.

\bibitem{techplayon2024cellreselection}
Techplayon, ``Cell reselection procedure,''
  \url{https://www.techplayon.com/cell-reselection/}, 2024, accessed:
  2025-07-30.

\bibitem{sirdc_lte_parameter_engineering_2025}
M.~Fadl, ``Lte parameter engineering guidelines,''
  \url{https://www.scribd.com/document/503090909/LTE-Parameter-Engineering-guidelines},
  2025, uploaded to Scribd; Accessed: 2025-07-30.

\bibitem{prasadnandkumar2025lte_ran_parameter_mapping}
P.~Nandkumar, ``{LTE RAN} {Ericsson-Huawei-Nokia} parameter mapping,''
  \url{https://pdfcoffee.com/lte-ran-ericsson-huawei-nokia-parameter-mapping-byprasadnandkumar-pdf-free.html},
  n.d., uploaded via PDFCoffee; Accessed: 2025-07-30.

\bibitem{ngmn2008_son_oam_requirements}
{NGMN Alliance}, ``Ngmn recommendation on self-organising networks (son) and
  operations \& maintenance (o\&m) requirements,''
  \url{https://www.ngmn.org/wp-content/uploads/NGMN_Recommendation_on_SON_and_O_M_Requirements.pdf},
  Next Generation Mobile Networks Alliance, Requirement Specification, 2008,
  accessed: Accessed: 2025-07-30.

\bibitem{osm2025}
{OpenStreetMap contributors}, ``{OpenStreetMap data retrieved via Overpass
  API},'' \url{https://overpass-api.de/api/interpreter}, 2025, data available
  under the Open Database License (ODbL).

\bibitem{2024-tnsm-multipoint}
S.~Schneider, H.~Karl, R.~Khalili, and A.~Hecker, ``Multi-agent deep
  reinforcement learning for coordinated multipoint in mobile networks,''
  \emph{{IEEE} Trans. Netw. Serv. Manag.}, vol.~21, no.~1, pp. 908--924, 2024.

\bibitem{2020-mobicom-icellspeed}
H.~Deng, Q.~Li, J.~Huang, and C.~Peng, ``{iCellSpeed:} increasing cellular data
  speed with device-assisted cell selection,'' in \emph{{ACM MobiCom}}, 2020.

\bibitem{2021-sigcomm-auric}
A.~Mahimkar, A.~Sivakumar, Z.~Ge, S.~Pathak, and K.~Biswas, ``Auric: using
  data-driven recommendation to automatically generate cellular
  configuration,'' in \emph{{ACM} {SIGCOMM}}, 2021.

\bibitem{2019-infocom-collaborative}
J.~Chuai, Z.~Chen, G.~Liu, X.~Guo, X.~Wang, X.~Liu, C.~Zhu, and F.~Shen, ``A
  collaborative learning based approach for parameter configuration of cellular
  networks,'' in \emph{{IEEE INFOCOM}}, 2019.

\bibitem{2022-imc-aurora}
A.~Mahimkar, Z.~Ge, X.~Liu, Y.~Shaqalle, Y.~Xiang, J.~Yates, S.~Pathak, and
  R.~Reichel, ``Aurora: conformity-based configuration recommendation to
  improve {LTE/5G} service,'' in \emph{{ACM IMC}}, 2022.

\bibitem{2023-mobicom-chroma}
C.~Ge, Z.~Ge, X.~Liu, A.~Mahimkar, Y.~Shaqalle, Y.~Xiang, and S.~Pathak,
  ``Chroma: Learning and using network contexts to reinforce performance
  improving configurations,'' in \emph{{ACM MobiCom}}, 2023.

\bibitem{2018-imc-mobility}
H.~Deng, C.~Peng, A.~Fida, J.~Meng, and Y.~C. Hu, ``Mobility support in
  cellular networks: {A} measurement study on its configurations and
  implications,'' in \emph{{ACM IMC}}, 2018.

\bibitem{2016-sigmetrics-mobility}
Y.~Li, H.~Deng, J.~Li, C.~Peng, and S.~Lu, ``Instability in distributed
  mobility management: Revisiting configuration management in 3g/4g mobile
  networks,'' in \emph{{ACM} {SIGMETRICS}}, 2016.

\bibitem{2020-sigcomm-5gmobility}
Y.~Li, Q.~Li, Z.~Zhang, G.~Baig, L.~Qiu, and S.~Lu, ``Beyond 5g: Reliable
  extreme mobility management,'' in \emph{{ACM} {SIGCOMM}}, 2020.

\bibitem{2025-infocom-handover}
M.~Kalntis, A.~Lutu, J.~A.~O. Iglesias, F.~A. Kuipers, and G.~Iosifidis,
  ``Smooth handovers via smoothed online learning,'' in \emph{{IEEE}
  {INFOCOM}}, 2025.

\bibitem{2019-csur-rlnet}
N.~C. Luong, D.~T. Hoang, S.~Gong, D.~Niyato, P.~Wang, Y.~Liang, and D.~I. Kim,
  ``Applications of deep reinforcement learning in communications and
  networking: {A} survey,'' \emph{{IEEE} Commun. Surv. Tutorials}, vol.~21,
  no.~4, pp. 3133--3174, 2019.

\bibitem{2017-sigcomm-pensieve}
H.~Mao, R.~Netravali, and M.~Alizadeh, ``Neural adaptive video streaming with
  pensieve,'' in \emph{{ACM} {SIGCOMM}}, 2017.

\bibitem{2024-infocom-rlcc}
L.~Giacomoni and G.~Parisis, ``Reinforcement learning-based congestion control:
  {A} systematic evaluation of fairness, efficiency and responsiveness,'' in
  \emph{{IEEE} {INFOCOM}}, 2024.

\bibitem{2024-infocom-fumes}
M.~Bl{\"{o}}cher, N.~Nedderhut, P.~Chuprikov, R.~Khalili, P.~Eugster, and
  L.~Wang, ``Train once apply anywhere: Effective scheduling for network
  function chains running on {FUMES},'' in \emph{{IEEE} {INFOCOM}}, 2024.

\bibitem{2021-pimrc-slicing}
H.~Zhou, M.~H.~M. Elsayed, and M.~Erol{-}Kantarci, ``{RAN} resource slicing in
  {5G} using multi-agent correlated {Q-Learning},'' in \emph{{IEEE} {PIMRC}},
  2021.

\end{thebibliography}
